\documentclass[aps,prd,twocolumn,superscriptaddress]{revtex4-1} 
\usepackage{fullpage, amsmath, xparse, amssymb, mathtools, graphicx, braket,xcolor,verbatim}
\usepackage{natbib}
\usepackage[colorlinks=true,linkcolor=blue,citecolor=blue]{hyperref}

\newcommand{\mk}[1]{{\textcolor{green}{#1}}}

\usepackage{multirow}
\usepackage{cancel}

\begin{document}
\title{Field moment expansion method for interacting Bosonic systems}
\author{Andrew Eberhardt}
\email{Corresponding author. \\ aeberhar@stanford.edu}
\affiliation{Kavli Institute for Particle Astrophysics and Cosmology, Menlo Park, 94025, California, USA}
\affiliation{Physics Department, Stanford University, Stanford, California, USA}
\affiliation{SLAC National Accelerator Laboratory}
\author{Michael Kopp}
\affiliation{Nordita,
KTH Royal Institute of Technology and Stockholm University,
Hannes Alfv\'ens v\"ag 12, SE-106 91 Stockholm, Sweden}
\author{Alvaro Zamora}
\affiliation{Kavli Institute for Particle Astrophysics and Cosmology, Menlo Park, 94025, California, USA}
\affiliation{Physics Department, Stanford University, Stanford, California, USA}
\affiliation{SLAC National Accelerator Laboratory}
\author{Tom Abel}
\affiliation{Kavli Institute for Particle Astrophysics and Cosmology, Menlo Park, 94025, California, USA}
\affiliation{Physics Department, Stanford University, Stanford, California, USA}
\affiliation{SLAC National Accelerator Laboratory}
\begin{abstract}

We introduce a numerical method and python package, \href{https://github.com/andillio/CHiMES}{CHiMES}, that simulates quantum systems initially well approximated by mean field theory using a second order extension of the classical field approach. We call this the field moment expansion method. In this way, we can accurately approximate the evolution of first and second field moments beyond where the mean field theory breaks down. This allows us to estimate the quantum breaktime of a classical approximation without any calculations external to the theory. We investigate the accuracy of the field moment expansion using a number of well studied quantum test problems. Interacting Bosonic systems similar to scalar field dark matter are chosen as test problems. We find that successful application of this method depends on two conditions: the quantum system must initially be well described by the classical theory, and that the growth of the higher order moments be hierarchical.

\end{abstract}

\maketitle

\section{Introduction}

Interacting many body bosonic systems describe a wide array of interesting phenomena. This includes Bose-Einstein condensates (BEC) \cite{Anderson1995, Davis1995}, electromagnetic radiation \cite{bial1977}, and scalar field dark matter (SFDM) \cite{Hu2000, mocz2019, Guth2015}. Their dynamical properties are often explored using a classical mean field theory (MFT) approximation, the Gross-Pitaevskii equations, or Schr\"odinger-Poisson equations in the case of SFDM \cite{Hertberg2016, Hu2000, Guth2015, Leggett2001, Carr2000, Baer2000, Minguzzi2004}.   

Numerically, MFT is preferable to an exact quantum field description which, for a system with $M$ interacting modes and total number of particles $n_{tot}$, would involve simulating a Hilbert space of dimensional $\sim n_{tot}^{M-2}$ \cite{Sikivie:2016enz}. For large $M$ or $n_{tot}$ an exact quantum treatment is infeasible. Rather than try and implement an exact quantum solver it is simpler to extend the classical theory using correction terms that capture quantum effects on the classical physics \cite{Alon2007, Prezhdo2000, 2020Bossman, Heller1976TimeDV, 1992Royer, sreedharan2020, Lentz2019}.

When occupation numbers are large and interactions weak the MFT is known to accurately describe the dynamics of these systems \cite{Leggett2001, bial1977,Gross, Pi, BECrev}. However, any interacting system with a nonlinearity will exhibit quantum corrections on some time scale \cite{Yurke1986, sreedharan2020, Lewenstein1996, Cabellero2008, Chakrabarty:2017fkd, chakrabarty2021, kopp2021nonclassicality}. The MFT, tracking only the mean value of the field operator, cannot account for these quantum terms \cite{Alon2007}. This means that effects like phase diffusion, quantum squeezing, and fragmentation inherently require a beyond MFT approach \cite{Alon2007, sreedharan2020, Yurke1986, Minguzzi2004}. The effect of these corrections on MFT is a current topic of interest \cite{Dvali_2018, Hertberg2016, Sikivie:2016enz, Dvali:2017eba, sreedharan2020, Chakrabarty:2017fkd, kopp2021nonclassicality, Alon2007, sreedharan2020, Lentz2019, Lentz2020}. This motivates the development of numerical methods which can capture beyond MFT physics \cite{Alon2007, Prezhdo2000, 2020Bossman, Heller1976TimeDV, 1992Royer, sreedharan2020, Lentz2019}.  We will refer to the time at which MFT can no longer accurately approximate the evolution of the underlying system as the ``quantum breaktime". 

The classical theory is generally achieved as a limit of the quantum field theory. When the expectation value of the field operator is large compared to the variance of the field operator it is sensible to replace the field operators in the quantum field theory with their expectation values \cite{Ball}. The expectation value of the quantum field is then called the classical field. Quantum coherent states, with parameters large compared to unity, satisfy this approximation criterion \cite{Ball}. Tracking higher order moments and their effects on the evolution of observables has been studied with success for position and momentum operators \cite{Heller1976TimeDV, Prezhdo2000}. Likewise, conceptually similar methods of expanding the field about its mean value have proven useful as corrections to MFT \cite{1992Royer, Georges_1991, Lentz2019, Polkovnikov_2010} as well as helping explore interesting physics \cite{GLIMM1976, 2020Bossman, TANAS1983351, Leggett2001}.

We apply these techniques to extend the MFT approximation to include terms proportional to higher order central moments and then integrate coupled differential equations governing the evolution of the the mean and these higher order moments. We will refer to this method as the ``field moment expansion" (FME). The main focus of this work will be to introduce a publicly available code repository which implements a solver which tracks both the mean field values and the second moments of the field. In this way we can simulate systems initially well described by MFT into regimes where quantum corrections become important.

There are two main benefits to this approach. First, prior to the quantum breaktime, FME produces a more accurate approximation of the quantum field expectation values than MFT. Secondly, FME provides an internal estimation of the breaktime, and can therefore estimate its own regime of validity. This internal assessment means no calculations external to the theory are necessary to estimate the quantum breaktime. This is in contrast to approximations of this timescale calculated by methods external to MFT, see for example \cite{Dvali_2018, Dvali:2013vxa, Dvali:2017eba}. Additionally, when compared with other mean field extensions \cite{Alon2007}, the FME scales as $\mathcal{O}(M^2 \, \log M)$, depending only on the grid size as opposed to the particle occupation numbers $n_p$.

The method is applicable to any interacting scalar field system assuming that a number of criteria are met. First, the initial correction terms must be subleading order. The method predictions cannot be trusted past the time when the quantum corrections become large. However, we will show that the FME is able to approximate the evolution of the quantum system longer than MFT. Second, the correction terms must grow hierarchically. For a term, $F$ of order $m$ written as a function of moments of order less than and equal to $m$, given $F^m(\text{moments of order $\le m$})$, the evolution the terms must satisfy $F^1 > F^2 > F^3 \dots$. Generally, these criteria will be met if the initial conditions are a coherent state with large mode occupation numbers. 

In this work we test the field moment expansion using two test problems that have been well studied in the literature\mk{,} for which exact quantum solutions are possible and that exhibit a breakdown of the MFT on some timescale \cite{Yurke1986, Hertberg2016, Sikivie:2016enz, Sikivie2012}. For each we show that the FME provides a more accurate solution until the quantum breaktime. Most importantly, we demonstrate that the method can be used in this case to accurately predict the quantum breaktime. 

The paper is organized as follows. In Section \ref{sec:Background}, we discuss background on interacting Bosonic systems, the MFT, and FME approximations. Section \ref{sec:numerics} explains our numerical implementation. In Section \ref{sec:testProbs}, we demonstrate that FME is accurate for a number of quantum test problems. Conclusions regarding the overall utility of these methods and future directions are presented in Section \ref{sec:conclusions}.

\section{Background} \label{sec:Background}
\subsection{Quantum description}

We start from the following Hamiltonian, used to describe non-relativistic scalar fields \cite{Hertberg2016, Sikivie:2016enz, Sikivie2012}. 

\begin{equation} \label{Ham}
    \hat H = \sum_j^M \omega_j \hat a_j^\dagger \hat a_j + \sum_{ijkl}^M \frac{\Lambda_{kl}^{ij}}{2} \hat a_k^\dagger \hat a_l^\dagger \hat a_i \hat a_j \, .
\end{equation}

Where the sums are performed over the $M$ system modes. For appropriately chosen $\Lambda$, $\omega$, and $M$ this Hamiltonian describes a wide range of physical systems. The first sum describes the kinetic energy of the system and the second the self interactions. $\hat a_i$ is the annihilation operator on mode $i$, which is defined by its commutation and action on number eigenstates. A number eigenstate is written as 

\begin{equation}
    \ket{\set{n}} = \ket{n_1, n_2, \dots, n_{M}} \, ,
\end{equation}

where $n_i$ describes the number of particles occupying the $i$th mode. Here we will take the modes to represent momentum eigenstates with momentum $p_i$. The number eigenstates form an orthonormal basis such that $\braket{n'_j|n_i} = \delta_{i \, j} \delta_{n \, n'}$. We can now describe the $\hat a$ operators as follows 

\begin{align}
    [\hat a_i, \hat a_j] &= 0 \, , \\
    [\hat a_i, \hat a^\dagger_j] &= \delta_{ij} \, , \\
    \hat a_j^\dagger \ket{n_j} &= (n_j + 1)^{1/2} \ket{n_j + 1} \, , \\
    \hat a_j \ket{n_j} &= n_j^{1/2} \ket{n_j - 1} \, , \\
    \hat{N}_j \ket{n_j} &\equiv \hat a_j^\dagger \hat a_j \ket{n_j} = n_j \ket{n_j}\, \, .
\end{align}

The annihilation operator can also be used to define the complex quantum field $\hat \psi(x)$, which is related to $\hat a$ by Fourier transform.

\begin{equation} \label{psi2a}
    \hat \psi(x) = \sum_i \hat a_i u^\dagger_i(x)
\end{equation}

where $u^\dagger_i(x)$ is the eigenstate of the momentum operator with eigenvalue $p_i$. The Heisenberg equation describes the dynamics of these operators. For an arbitrary operator $\hat A$ with time independent Hamiltonian the equation of motion is written 

\begin{equation} \label{HeisenbergEq}
    \partial_t \hat A = \frac{i}{\hbar} [\hat H, \hat A] \, .
\end{equation}

Hereafter we set $\hbar \equiv 1$. We can now solve for the evolution of our field operators,

\begin{align} \label{eqnMotion}
    \partial_t \hat a_p &= i [\hat H, \hat a_p] = -i\left[ \omega_p \hat a_p + \sum_{ijl} \Lambda^{ij}_{pl} \hat a_l^\dagger \hat a_i \hat a_j \right] \, .
\end{align}

In this work we will be taking the constants $\Lambda^{ij}_{pl}$ to be of the following form

\begin{equation} 
    \Lambda^{ij}_{pl} = \left( \frac{C}{2(p_p - p_i)^2} + \frac{C}{2(p_p - p_j)^2} + \Lambda_0 \right) \delta^{ij}_{pl} \,,
\end{equation}
where the constant $C$ describes a long range $r^{-1}$ potential and $\Lambda_0$ characterizes the strength of contact interactions. $\delta^{ij}_{pl}$ is the Kroneker delta. If $C = \frac{-4 \pi G m^2}{L}$ and $\Lambda_0 = 0$, then taking a Fourier transform of equation \eqref{eqnMotion}  yields the familiar second quantized Schr\"odinger--Poisson equations. Where $G$ is the gravitational constant, $L$ is the volume of the box for which the quantum field is periodic, and $m$ is the mass of the field. Here we are working in 1-D.

\begin{align}
    \partial_t \hat \psi(x) &= -i \left[ \frac{-\nabla^2}{2m}  + m \hat V(x)  \right] \hat \psi(x) \, , \\ 
    \nabla^2 \hat V &= 4 \pi G m \, \hat \psi^\dagger(x) \hat \psi(x) \, .
\end{align} 
Here we have started with a complex quantum field operator. However, it is possible to derive this set of equations as a non-relativistic and weak gravity limit of the real scalar Klein Gordon field. This can be done following the derivation in \cite{Suarez2015, Lentz2019}. These limits need to be kept in mind when determining where this set of equations and approximations of it are valid.

While the above analysis is true for an arbitrary quantum state, within the stated limits, it is useful to define an initial quantum state for which the mean field theory starts as an accurate approximation of the quantum field theory. The ``most classical" state is the coherent state, for which MFT is initially exact, parameterized by the complex vector $\Vec{z} \in \mathbb{C}^M$, which can be written as a sum of number eigenstates as 

\begin{equation} \label{coherentStates}
    \ket{\Vec{z}} = \bigotimes_{i=1}^M \exp \left[ -\frac{|z_i|^2}{2} \right] \sum_{n_i=0}^\infty \frac{ z_i^{n_i}}{\sqrt{n_i!}} \ket{n_i} \, .
\end{equation}

When representing a coherent state numerically we truncate the above sum when the square norm of $\braket{\vec z | \vec z} \ge .995$.

A coherent state is thought to describe the initial state of the axion field if produced via the misalignment mechanism \cite{ABBOTT1983133, Preskill:1982}.

This state implies that a measurement of the particle number in the mode $i$ would be Poisson distributed with expectation value $|z_i|^2$. This state is special because it has the property that the expectation value any normally ordered operator composed of $\hat a$ and $\hat a^\dagger$ with respect to this state is given by simply replacing the operators with the parameter $z$, i.e. 

\begin{equation} \label{normalOrder}
    \braket{\vec z | \, f(\set{\hat a^\dagger}) \, g( \set{\hat a}) \, | \vec z } = f(\vec{z}^\dagger) \, g(\vec{z}) \, .
\end{equation}

This will be important when deriving the mean field theory. 

\subsection{Mean field approximation}
The mean field is simply the expectation value of the field operator, we define a mean field in position and momentum space respectively as

\begin{align}
    \psi(x) &\equiv \braket{\hat \psi(x)} \, , \\
    a_i &\equiv \braket{\hat a_i} \, .
\end{align}

The higher order moments can then be calculated from the mean field operators. Occupation numbers, $N^{cl}$ are given as the amplitude of the field operators, e.g. 

\begin{equation} \label{clApprox}
    N_i^{cl} = |a_i|^2
\end{equation}

The mean field theory is attained simply by taking an expectation value of the equations of motion and then approximating the operators by their expectation value. Let us say that the operator $\hat A$ evolves according to the following equation of motion

\begin{equation} \label{opFun}
    \partial_t \hat A = f(\hat{A}) \, .
\end{equation}

And that $\hat A$ corresponds to some dynamic observable of the system. If the expectation value of $\hat A$ is large compared to its root variance than we can make the following approximation

\begin{equation} \label{mft}
    \partial_t \braket{\hat A} = \braket{f(\hat A)} \approx f(\braket{\hat A}) \,.
\end{equation}

This approximation is one way to transition to a mean field theory description. It is identical to Ehrenfest theorem if we replaced the position and momentum operators with field operators. 

This statement that the mean field theory is accurate at a time $T$ implies the following conditions

\begin{equation} \label{mfcondition}
    \mathrm{E}[\hat A(T)] \gg \sqrt{\mathrm{Var}[\hat A(T)]} \, .
\end{equation}

Note this is a condition both on the evolution of the operator $\hat A$ and the quantum state that the expectation is taken with respect to.

For the mean field approximation to hold we need only that the approximation in equation \eqref{mft} remain accurate on the time scale of the evolution. We will make the requirements more precise in the next section, but from here we can see qualitatively why large occupation number tends to motivate the mean field theory approximation. 

Let us assume that $\hat A = \set{\hat a, \hat a^\dagger, \dots}$ is the set of operators generated by field operators $\hat a$ and $\hat a^\dagger$, as will be the case in the next sections. It is always possible to write the right hand side of equation \eqref{opFun} in terms of normally ordered operators. This means that if we are in a coherent state by equation \eqref{normalOrder} the mean field approximation is an equality so long as we remain in a coherent state. This is true regardless of expected occupation number, however, if we are only in an approximately coherent state with $|z|^2 = n \gg 1$ then we can expect the variance in our field operators to be approximately governed by a Poisson distribution; i.e. $\mathrm{E}[\hat N] \sim \mathrm{Var}[\hat N] \sim |z|^2 = n$, this means that the fractional deviation in the expectation value will go as $\sqrt{\mathrm{Var}[\hat N]}/\mathrm{E}[\hat N] \sim 1/\sqrt{N}$. This is easily made into an estimate of the fractional field variance by recalling that $\hat N = \hat a^\dagger \hat a$. So the fractional deviation in the field values is small for large occupation numbers $n \gg 1$. 

Large occupation numbers is not enough on its own to ensure an accurate MFT description. The mean field theory will be accurate if the occupation numbers be large but also that the quantum state remained approximately coherent, or equivalently, that the distribution of the number operator has Poisson distributed expectation and variance proportional to some power of $n$. But the actual condition that needs to be satisfied is equation \eqref{mfcondition}.

It is easy to imagine a state for which this assumption is not met. A number eigenstate, $\ket{\set{n}}$, for example has field expectation $\braket{\set{n}| \, \hat a \, | \set{n}} = 0$ even when $n \gg 1$. Therefore, it does not satisfy the condition in equation \eqref{mfcondition} and will not be well described by a single classical field even for large $n$. It was demonstrated in \cite{Sikivie:2016enz, chakrabarty2021} that a number eigenstate did not approach a single classical field description even at large occupation number. It was then shown in \cite{Hertberg2016} that this state could be approximated by an ensemble of classical fields with ensembled expectation $0$ and amplitude $n$.

It is important to note that large occupation number is only a proxy for the accuracy of the mean field theory. For example, a coherent state evolved by the free particle Hamiltonian will always be perfectly described by classical field theory even as $\mathrm{E}[\hat N] \ll 1$. Conversely, a number eigenstate will never be well described by a single classical field even as $\mathrm{E}[\hat N] \gg 1$. Going forward we will phrase our estimation of the accuracy of the classical field theory in terms of the condition described in equation \eqref{mfcondition} for the field operator $\hat a$.

It should also be noted that it is possible to reproduce the quantum evolution of the number operator without reproducing the evolution of the field itself. For example, a so-called field number state in the large $N$ limit approaches the classical evolution of the mode occupations, however, it has vanishing field expectation regardless of $N$. For states like these, equation \eqref{mfcondition} should be expressed in terms of the number operator.

\subsection{Quantum corrections}

Systems initially well described by the mean field theory will eventually diverge from this description on some time scale if the Hamiltonian is non-Harmonic. The specific causes of this are of interest in the literature \cite{Lewenstein1996, sreedharan2020, Cabellero2008, Yurke1986, kopp2021nonclassicality} but for our purposes we can think of them generally as a delocalization in phase space. In this section we will show an example of how deviation from the classical field theory occurs, and discuss a way to parameterize it.

Specifically that the variance in the field operators becomes of order the expectation value violating the condition in equation \eqref{mfcondition}. We will clarify this point using the following example.

Consider the toy Hamiltonian on one mode

\begin{equation} \label{pqHam}
    \hat H = \frac{\hat p^2}{2m} +  \lambda_L \hat q^2 + \lambda_{NL} \hat q^4 \, .
\end{equation}

Where $p \equiv -i\hbar \nabla_q$ The first two terms define a harmonic oscillator and the last term some non-linearity. The set of operators that we are interested in are the position and momentum operators $\hat A = \set{\hat q, \hat p}$. Note that this Hamiltonian can be recast in terms of the $\hat a$ operator using the relation

\begin{equation} \label{a2pq}
    \hat a = \frac{1}{\sqrt{2}} (\hat q + i\hat p) \, ,
\end{equation}

but this does not change the analysis. 

The classical equations of motion can be found using equations \eqref{HeisenbergEq} and \eqref{pqHam} as well as the commutation relation $[q,p] = i \hbar$ and then applying Ehrenfest's theorem. For the classical variables $p$, $q$, related to the classical field by $a = \frac{1}{\sqrt{2}}(q + ip)$ (recall there is only a single mode), we obtain the following equations of motion

\begin{align}
    \partial_t p &= -2 \,  \lambda_L \, q -4 \, \lambda_{NL} \, q^3 \\
    \partial_t q &= p
\end{align}

With $(q,p)|_{t=0} = (q_i, 0)$. We use a symplectic leap-frog integrator to solve the classical equations of motion. 

Our initial wavefunction will be Gaussian, which is initially well localized in $q$,

\begin{equation}
   \braket{q | \phi } = \sqrt{\frac{1}{\sqrt{2 \pi \sigma}}} \exp\left(-\frac{(q-q_i)^2}{4 \sigma^2}\right)
\end{equation}

The quantum equations can be solved by integrating Schr\"odinger's equation

\begin{equation}
    \partial_t \ket{\phi} = -i \hat H \, \ket{\phi} \, .
\end{equation}

We use a symplectic spectral leap-frog integrator to perform the integration. 

Let us also parameterize the inequality in equation \eqref{mfcondition} by defining the following quantity $Q_q$

\begin{equation}
    Q_q \equiv \frac{\braket{\hat q^2} - \braket{\hat q}^2}{q_i^2} \, .
\end{equation}

This can be used to parameterize the quantum theories deviation from the classical theory because the nonlinearity is spatial. It should be noted that this is not the only parameter that can be constructed with this property. 

\begin{figure*} 
	\includegraphics[width = .97\textwidth]{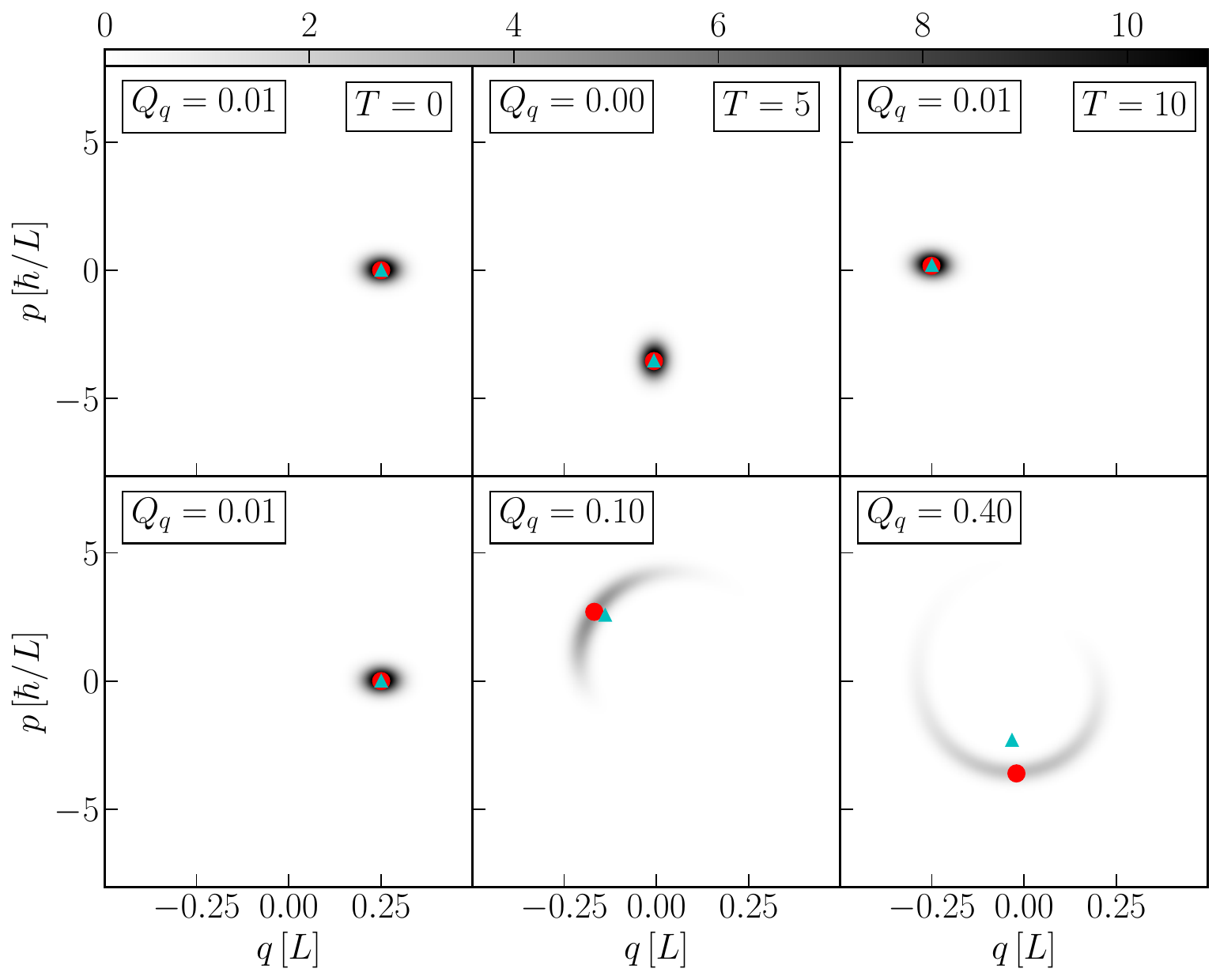}
	\caption{Here we show how quantum corrections cause deviations from the mean field theory. Plotted are the Husimi functions for two different non-linear models at three different times. The top row corresponds to the weakly nonlinear case with $\lambda_{NL} = 0.073$, and the bottom row to the strongly nonlinear case with $\lambda_{NL} = 73$. Each column corresponds to a different time, $T$ in the evolution. The red dot in each panel indicates the solution solved obtained using the classical mean field theory, $a$. The cyan triangle in each panel represents the actual mean of the $\hat a$ operator $\braket{\hat a}$. At all times the weakly nonlinear model closely adheres to the classical solution. The strongly nonlinear case has the classical field theory start as a good approximation but strongly diverge over the course of the evolution. }
	\label{fig:quantumErrors}
\end{figure*}

In figure \ref{fig:quantumErrors} we track a quantum phase space analog, the Husimi function \cite{Ball}, of the wavefunction, the classical mean field theory approximation of $\mathrm{E}[\hat a]$, the exact quantum value of $E[\hat a]$ for two different strength nonlinearities. In both cases the initial spread of the wavefunction is $\sigma = 0.025$, the initial location $q_i = 0.25$, $\hbar = 0.01$, $\lambda_L = 100$. This means that our parameter $Q_q = 0.01$. Note also that equation \eqref{a2pq} implies that the occupation number here is $n = q_i^2 = 0.0625$. Clearly we are not in the high occupation regime. The top row and bottoms rows have $\lambda_{NL} = 0.073$ and $\lambda_{NL} = 73$, representing the weakly nonlinear and strongly nonlinear respectively. 

For the nonlinear case we see in the second panel that squeezing and phase diffusion have caused the Husimi function to be poorly localized around the classical solution, however, at this time the Husimi function is still well approximated by a squeezed Gaussian. At this time the classical and quantum solutions begin to diverge. This is approximately the quantum breaktime, where the $Q_q$ parameter is starting to approach $\mathcal{O}(1)$. In the rightmost panel of the nonlinear evolution we see that phase diffusion has removed most of the information about the phase of the field. The classical and quantum solutions now deviate by an $\mathcal{O}(1)$ fraction. 

Throughout this work we will be interested in functions of the field operators and their Fourier transforms. We will define the quantum breaktime as we have done in this section, by first defining a quantity that parameterizes the deviation from the classical theory. The following definition will be used moving forward

\begin{equation} \label{eqQ}
    Q \equiv \sum_i^M \frac{\braket{\delta \hat a_i^\dagger \delta \hat a_i}}{n_{tot}} \, .
\end{equation}

Where $n_{tot} \equiv \sum_i^M n_i$ and $\delta \hat a \equiv \hat{a} - \braket{\hat a}$.

When $Q$ is small a single classical field can accurately capture both the first and second moments of the field operator and the classical approximation in equation \eqref{clApprox} is valid. When this parameter ceases being small, such a description inaccurately capture both the field and occupation number expectations. Therefore, we will use $Q$ to define a quantum breaktime, $t_{br}$ condition as when $Q = .15$, i.e.

\begin{equation} \label{breakTime}
    Q(t_{br}) \equiv 0.15 \, .
\end{equation}

Note that this is not intended to be the unique usable definition of the quantum breaktime. It is only intended to allow us to quantify when the assumption in equation \eqref{clApprox} breaks down. The specific value $0.15$ is arbitrary and the scaling of the breaktime with occupation number should be relatively insensitive to the specific choice of $Q(t_{br})$. We choose this specific value because it indicates that the correction terms are becoming the same order as the leading order classical terms and analysis of our test problems indicate that the field moment solver reliably reproduces quantum results when $Q<0.15$.

\subsection{Field moment expansion}

For well behaved probability distributions with well defined moments, we can uniquely identify the distribution by its moments. This is true for functions of quantum mechanical operators. Consider a set of operators $\hat A$ and function of this set $f(\hat A)$. Let us assume we can write $f$ as a sum of products of normally ordered generators of $\hat A$, and that there exists some integer $R$ for which every term contains $R$ or fewer elements of the generator of $\hat A$. Let us also assume that the $i$th moment of each element of $\hat A$ is well defined for $i \le R$. We can then write the expectation value of $f$ as a sum of terms weighted by central moments as \begin{widetext}

\begin{equation} \label{formalExp}
    \braket{f(\hat A)} = f(\braket{\hat A}) 
    + \sum_{j=2}^R \frac{1}{j!} \left[ \prod_{k=1}^{j} \left( \sum_{\hat a \in \hat A} \hat \delta_{\hat a}   \right) \right] f( \braket{\hat A} ) \, .
\end{equation}

Where $\braket{\hat A} \equiv \set{\braket{a_1}, \braket{a_2}, \dots }$ is the set of the expectation values of the elements of the set $\hat A$, and the product of the $\hat \delta_{\hat a}$ operators is defined as 

\begin{equation}
    \hat \delta_{\hat a_1} \hat \delta_{\hat a_2}  \dots \equiv \braket{\delta \hat a_1 \, \delta \hat a_2 \, \dots } \frac{\partial}{\partial \braket{\hat a_1} }  \frac{\partial}{\partial \braket{\hat a_2} } \dots \, .
\end{equation}

Where the operators $\delta \hat a$ are normally ordered. 

Consider the Hamiltonian in equation \eqref{Ham}. We can solve for the equation of motion for the $\hat a$ operator using equation \eqref{HeisenbergEq} giving

\begin{align} \label{eqnMotion}
    \partial_t \hat a_p &= i [\hat H, \hat a_p] = -i\left[ \omega_p \hat a_p + \sum_{ijl} \Lambda^{ij}_{pl} \hat a_l^\dagger \hat a_i \hat a_j \right] \, .
\end{align}

We see that the highest order operator in the equation of motion is third order in $\hat a$ and $\hat a^\dagger$. Replacing the $\hat a$ and $\hat a^\dagger$ operators with their expectation values in this equation gives the classical field theory given by the following inequality

\begin{align}
    \partial_t \braket{\hat a_p} &= i \braket{[\hat H, \hat a_p]} \nonumber \\ &= -i\left[ \omega_p \braket{\hat a_p} + \sum_{ijl} \Lambda^{ij}_{pl} \braket{\hat a_l^\dagger \hat a_i \hat a_j} \right] \label{exactEOM} \\
    &\approx -i\left[ \omega_p \braket{\hat a_p} + \sum_{ijl} \Lambda^{ij}_{pl} \braket{\hat a_l^\dagger} \braket{\hat a_i} \braket{\hat a_j} \right] \, .
\end{align}

We can rewrite equation \eqref{exactEOM} the form of equation \eqref{formalExp}. 
\begin{subequations} \label{expandedEOM}
\begin{align}
    \partial_t &\braket{\hat a_p} = i \braket{[\hat H, \hat a_p]} \nonumber \\ &= -i\left[ \omega_p \braket{\hat a_p} + \sum_{ijl} \Lambda^{ij}_{pl} \braket{\hat a_l^\dagger \hat a_i \hat a_j} \right] \nonumber \\
    &= -i\left[ \omega_p \braket{\hat a_p} + \sum_{ijl} \Lambda^{ij}_{pl} \left( \braket{\hat a_l^\dagger} \braket{\hat a_i} \braket{\hat a_j} \right. \right.  \label{firstOrder} \\
    &+ \braket{\delta \hat a_i \delta \hat a_{j}} \braket{\hat a^\dagger_l} + \braket{\delta \hat a^\dagger_l \delta \hat a_{i}} \braket{\hat a_j} + \braket{\delta \hat a^\dagger_l \delta \hat a_{j}} \braket{\hat a_i}  \label{secondOrder} \\
    &+ \left. \left.\braket{\delta \hat a^\dagger_l \delta \hat a_{i} \delta \hat a_{j}} \right) \right] \, . \label{thirdOrder} \\
    &\approx -i\left[ \omega_p \braket{\hat a_p} + \sum_{ijl} \Lambda^{ij}_{pl} \left( \braket{\hat a_l^\dagger} \braket{\hat a_i} \braket{\hat a_j}  + \braket{\delta \hat a_i \delta \hat a_{j}} \braket{\hat a^\dagger_l} + \braket{\delta \hat a^\dagger_l \delta \hat a_{i}} \braket{\hat a_j} + \braket{\delta \hat a^\dagger_l \delta \hat a_{j}} \braket{\hat a_i} \, \right) \right] . \label{FME_EOM}
\end{align}
\end{subequations}
\end{widetext}

We see in this form that the classical equations of motion are given by equations \eqref{firstOrder}, then \eqref{secondOrder} and \eqref{thirdOrder} act as ``quantum" corrections to the mean field theory. We can also see now the manner in which our classicality condition in equation \eqref{mfcondition} is technically imprecise. What is actually required is that the terms in \eqref{secondOrder} and \eqref{thirdOrder} remain small compared to the terms in \eqref{firstOrder}. However, it is important to keep in mind that the accuracy of the mean field theory is not intrinsically a function of occupation number but instead a property of how the moments of the field operators compare. 

The condition in equation \eqref{mfcondition} comes about by assuming that the moments grow hierarchically, that is that the first order terms in \eqref{firstOrder} start out largest and that the terms in \eqref{secondOrder} grow faster than the terms in \eqref{thirdOrder} and so on. If we then also assume that the second order central moments in \eqref{secondOrder} are all approximately the same order, and that the dynamics are approximately number conserving, we see that taking a ratio of the first and second order terms gives us the parameter $Q$ and asserting that $Q \ll 1$ is equivalent then to equation \eqref{mfcondition}. 

In order to integrate equation \eqref{expandedEOM} we couple the evolution of the field operator $\hat a$ to the evolution of the central moments. If we assume that the central moments grow hierarchically and we are interested in evolution only until the classical field theory breaks we can truncate equation \eqref{expandedEOM} at the second order terms, this has the benefit of better computational scaling. The equations of motion for the higher order moments can be found using equation \eqref{HeisenbergEq} and then expanding to second order using equation \eqref{formalExp}. We will start with the $\mathrm{Cov}[\hat a_i, \hat a_j]$ operator which has the following equation of motion.

\begin{widetext}

\begin{align} \label{daa_dt}
    \partial_t \braket{\delta \hat a_i \delta \hat a_j} &\approx -i \left[(\omega_i+\omega_j)\braket{ \delta \hat a_i  \delta \hat a_j} + \sum_{kp} \Lambda^{kp}_{ji} \braket{\hat a_k} \braket{\hat a_p}  \right. \\
    &\left. + \sum_{kpl} \Lambda^{kp}_{ji} \left(  \braket{ \delta \hat a_l \delta \hat a_j} \braket{\hat a^\dagger_k} \braket{\hat a_p} + \braket{ \delta \hat a_j  \delta \hat a_p} \braket{\hat a^\dagger_k} \braket{\hat a_l} + \braket{\delta \hat a_k^\dagger \delta \hat a_j} \braket{\hat a_l} \braket{\hat a_p} \right) + (i \leftrightarrow j) \right]  \, . \nonumber
\end{align}

This equation can be broken down into three types of terms. The first term and the second line of terms are the kinetic and potential terms respectively. The terms are proportional to the covariance operator. The second term on the second line is proportional both to the potential energy and $[\hat a, \hat a^\dagger]$. This term guarantees that the covariance operator will grow even if initially zero so long as their is some nonlinearity in the Hamiltonian. 

We can solve for the $\mathrm{Cov}[\hat a^\dagger_i, \hat a_j]$ operator in the same manner. 

\begin{align} \label{dba_dt}
    \partial_t \braket{\delta \hat a_i^\dagger \delta \hat a_j} &\approx i \left[ (\omega_i - \omega_j)\braket{\delta \hat a_i^\dagger \delta \hat a_j} +  \right.  \\
    &\left.  + \sum_{kpl}  \Lambda^{kp}_{ji}  \left( \braket{\delta \hat a_j \delta \hat a_k} \braket{\hat a^\dagger_p} \braket{\hat a^\dagger_l} + \braket{\delta \hat a^\dagger_p \delta \hat a_j} \braket{\hat a^\dagger_l} \braket{\hat a_k} + \braket{\delta \hat a^\dagger_l \delta \hat a_l} \braket{\hat a^\dagger_p} \braket{\hat a_k} \right) + (c.c., \, i \leftrightarrow j) \right] \, . \nonumber
\end{align}

Which has the same structure as the previous equation without the term proportional to the commutation operator. $c.c$ indicates complex conjugate.

\end{widetext}

The term in equation \eqref{daa_dt} proportional to the the commutation between $\hat a$ and $\hat a^\dagger$ gives us a qualitative sense of how large occupation number implies classicality. If the quantum state is approximately coherent then these second order central moments are near $0$ by equation \eqref{normalOrder}. On a timescale $\sim \mathcal{O}(\Lambda^{-1})$ the second order central moments will have grown by a factor $[ \hat a, \hat a^\dagger] = 1$. In the large occupation number limit  $[ \hat a, \hat a^\dagger] = 1 \ll 1$ meaning our lowest order quantum corrections contribute vanishingly to the evolution of the mean field.

\subsection{Penrose-Onsager criterion}

When the Penrose-Onsager (PO) criterion \cite{Penrose1956} is satisfied we can write

\begin{equation} \label{POcriterion}
    \braket{\hat a^\dagger_i \hat a_j} = \vec{z}_i^\dagger \vec z_j \, .
\end{equation}

That is that the expectation values of the second field moment matrix $M_{ij} \equiv \braket{\hat a^\dagger_i \hat a_j}$ can be written as an outer product of a single vector $\vec z$ with its complex conjugate.

When the PO criterion is satisfied $\hat M_{ij}$ contains a single nonzero eigenvalue, called the principal eigenvalue, equal to the square norm of $\vec z$, i.e. $\lambda_{p} = \sum_i |z_i|^2$. Where $\vec z / \sqrt{\sum_i|z_i|^2}$ is the corresponding principal eigenvector, $\vec \xi_p$. This vector is not a priori equal to the classical field but when the classical field adequately describes the system we expect the PO criterion to be satisfied. 

Note that $\mathrm{Tr}[\hat M_{ij}] = n_{tot}$ and,  therefore, in number preserving systems the trace of $\hat M_{ij}$ is a conserved quantity,. When the system is well described by the classical theory we expect that the principal eigenvalue is very close to $n_{tot}$, more specifically we expect \cite{Leggett2001}

\begin{equation}\label{PO_param}
    \frac{\lambda_p}{n_{tot} } - 1 \ll 1 \, .
\end{equation}

Because the FME tracks both second central moment $\braket{\delta \hat a_i^\dagger \delta \hat a_j}$ and $\braket{\hat a_i}$ we can use this method to approximate $\hat M_{ij}$ as 

\begin{equation}
    \hat M_{ij}^{\mathrm{FME}} = \braket{\delta \hat a_i^\dagger \delta \hat a_j}^{\mathrm{FME}} + (\braket{\hat a_i}^\dagger \braket{\hat a_j})^{\mathrm{FME}} \, .
\end{equation}

It is important to note that while both $Q$ and equation \eqref{PO_param} parameterize the deviation from the classical theory the two are technically distinct in the following way: 

When $Q \not\ll 1$ it implies that the classical approximation in equation \eqref{clApprox} is breaking down, i.e. $Q \sim 1\,, \rightarrow |\braket{\hat a}|^2 \ne \braket{\hat a^\dagger \hat a}$. Note that this is not a useful parameterization for states which track the evolution of the mode occupations but not the field values themselves. On the other hand, $\lambda_p/n_{tot} - 1 \not\ll 1$ implies that $\hat M_{ij}$ cannot be described by a single eigenvector. Neither explicitly implies that the classical field poorly approximates occupation numbers, but both can be used to give an approximate sense of how closely a system is adhering to the classical field theory.

\section{Numerical implementation} \label{sec:numerics}

The full code repository for the simulation and data analyses of the classical and expanded field theories performed here is publicly available at \href{https://github.com/andillio/CHiMES}{https://github.com/andillio/CHiMES}. 

\subsection{Mean field theory}

The evolution of the mean field, $a^{cl}$, is solved by integrating the classical field equations of motion given

\begin{align}
    \partial_t a^{cl}_p = -i\left[ \omega_p a^{cl}_p + \sum_{ijl} \Lambda^{kp}_{ji} a^{cl\dagger}_l a^{cl}_i a^{cl}_j \right] \, .
\end{align}
The initial conditions are chosen such that the field values correspond to a coherent state with parameter $\vec z = \vec a^{cl}$. Note that this implies the initial square amplitudes of the classical field give the mode occupation number expectations, i.e.

\begin{align}
    |a^{cl}_p|^2 \biggr\rvert_{t=0} = E[\hat N_p]
\end{align}

We use a fourth order Runga-Kutta update scheme to update the field \cite{Runge1895, Kutta}. The update function for the field at mode $p$ is given 

\begin{align}
    F(a_p) =  a_p \left( 1 - i \, \omega_p \, \Delta t \right) - i\, f(a)_p\,\Delta t \, ,
\end{align}

and the update scheme is then 

\begin{enumerate}
    \item $k_1 = F\left(a^{cl}_p(t)\right)$ 
    \item $k_2 = F\left( a^{cl}_p(t) + k_1/2 \right)$
    \item $k_3 = F\left( a^{cl}_p(t) + k_2/2 \right)$
    \item $k_4 = F\left( a^{cl}_p(t) + k_3 \right)$
    \item $a^{cl}_p(t + \Delta t) = \frac{1}{6} \left( k_1 + 2k_2 + 2 k_3 + k_4 \right)$
\end{enumerate}

which takes the field at a time $t$, $a^{cl}(t)$, to a time $t + \Delta t$, $a^{cl}(t + \Delta t)$. The function $f(a)$ defines the potential term and is given as follows 
\begin{align}
    &f(a)_p = \mathcal{F}\left[ V(x) \, \psi(x) \right]_p \, , \\
    &V(x) = \mathcal{F}^{-1}\left[ \mathcal{F}\left[ \psi^\dagger(y) \psi(y) \right]_i \left( \frac{C}{k_i^2} + \Lambda_0 \right)  \right](x) \, . \nonumber
\end{align}
Where $\mathcal{F}$ and $\mathcal{F}^{-1}$ define the Fourier transform and inverse Fourier transform respectively. $\psi(x) = \sum_i a^{cl}_i u^\dagger_i(x)$, as in equation \eqref{psi2a}, i.e. $\psi$ is the inverse Fourier transform of $a^{cl}$. In these simulations we use non-periodic boundary conditions. This is achieved by padding the ends of the arguments of the Fourier transform with $M/2$ zeros. Meaning our discrete Fourier transforms are defined as follows
\begin{subequations} \label{paddFourier}
\begin{align}
    \mathcal{F}[\psi(x)]_p &= \sum_{x=-M/2}^{M + M/2} \tilde \psi(x) \, u_p(x) \, , \\
    \mathcal{F}^{-1}[a_p](x) &= \sum_{p=-M/2}^{M + M/2} \tilde a_p \, u^\dagger_p(x) \, , 
\end{align}
\end{subequations}
and the padded fields are given
\begin{subequations}
\begin{align}
\tilde \psi(x) &= \begin{cases}
\psi(x) &\text{$0 \le x \le M$} \\
0 &\text{else}
\end{cases} \, , \\
\tilde a_p &= \begin{cases}
a_p &\text{$0 \le p \le M$} \\
0 &\text{else}
\end{cases} \, .
\end{align}
\end{subequations}
The unpadded fields can be recovered by looking only at the modes $\in[0,M]$.

\subsection{Field moment expansion}

The evolution of the field moments is solved by integrating the coupled equations \eqref{FME_EOM}, \eqref{daa_dt}, and \eqref{dba_dt}. The initial conditions are chosen such that the initial values of the moments correspond to those of a coherent state with parameter $\vec z$. Meaning
\begin{align}
 \braket{\hat a_p}^{\mathrm{FME}} \biggr\rvert_{t=0} &= z_p \, ,\\ 
 \braket{\delta \hat a_i \delta \hat a_j}^{\mathrm{FME}} \biggr\rvert_{t=0} &= 0 \, , \\
 \braket{\delta \hat a^\dagger_i \delta \hat a_j}^{\mathrm{FME}} \biggr\rvert_{t=0} &= 0 \, .
\end{align}
\begin{widetext}
This solver uses the following update functions

\begin{align}
    F^{\mathrm{FME}}_a(a_p, \, \delta A_{ij}, \, \delta B_{ij}) &= a_p \left( 1 - i \, \omega_p \, \Delta t \right) - i\,\Delta t \,\left( \, f(a^1)_p + g_1(a,\, \delta A_{ij})_p + g_2(a,\, \delta B_{ij})_p \, \right) \, , \\
    F^{\mathrm{FME}}_{aa}(a_p, \, \delta A_{ij}, \, \delta B_{ij}) &= \delta A_{ij} \left( 1 - i \, (\omega_i + \omega_j) \Delta t \right) - i \Delta t \left( h(a)_{ij} + g_3(a,\, \delta A_{ij})_{ij} + g_4(a,\, \delta B_{ij})_{ij} \right) \\
    F^{\mathrm{FME}}_{ba}(a_p, \, \delta A_{ij}, \, \delta B_{ij}) &= \delta B_{ij} \left( 1 + i \, (\omega_i - \omega_j) \Delta t \right) - i \Delta t \left(g_5(a,\, \delta A_{ij})_{ij} + g_6(a,\, \delta B_{ij})_{ij} \right)
\end{align}

We use the following Runga-Kutta integration scheme to update the field moments 

\begin{enumerate}
    \item $k_1^a = F^{\mathrm{FME}}_a\left(\braket{\hat a_p}^{\mathrm{FME}}(t), \, \braket{\delta \hat a_i \delta \hat a_j}^{\mathrm{FME}}(t), \, \braket{\delta \hat a^\dagger_i \delta \hat a_j}^{\mathrm{FME}}(t) \right)$
    \item $k_1^{aa} = F^{\mathrm{FME}}_{aa}\left(\braket{\hat a_p}^{\mathrm{FME}}(t), \, \braket{\delta \hat a_i \delta \hat a_j}^{\mathrm{FME}}(t), \, \braket{\delta \hat a^\dagger_i \delta \hat a_j}^{\mathrm{FME}}(t) \right)$
    \item $k_1^{ba} = F^{\mathrm{FME}}_{ba}\left(\braket{\hat a_p}^{\mathrm{FME}}(t), \, \braket{\delta \hat a_i \delta \hat a_j}^{\mathrm{FME}}(t), \, \braket{\delta \hat a^\dagger_i \delta \hat a_j}^{\mathrm{FME}}(t) \right)$
    \item $k_2^a = F^{\mathrm{FME}}_a\left(\braket{\hat a_p}^{\mathrm{FME}}(t) + k_1^a/2, \, \braket{\delta \hat a_i \delta \hat a_j}^{\mathrm{FME}}(t)+ k_1^{aa}/2, \, \braket{\delta \hat a^\dagger_i \delta \hat a_j}^{\mathrm{FME}}(t) + k_1^{ba}/2 \right)$
    \item $k_2^{aa} = F^{\mathrm{FME}}_{aa}\left(\braket{\hat a_p}^{\mathrm{FME}}(t) + k_1^a/2, \, \braket{\delta \hat a_i \delta \hat a_j}^{\mathrm{FME}}(t)+ k_1^{aa}/2, \, \braket{\delta \hat a^\dagger_i \delta \hat a_j}^{\mathrm{FME}}(t) + k_1^{ba}/2 \right)$
    \item $k_2^{ba} = F^{\mathrm{FME}}_{ba}\left(\braket{\hat a_p}^{\mathrm{FME}}(t) + k_1^a/2, \, \braket{\delta \hat a_i \delta \hat a_j}^{\mathrm{FME}}(t)+ k_1^{aa}/2, \, \braket{\delta \hat a^\dagger_i \delta \hat a_j}^{\mathrm{FME}}(t) + k_1^{ba}/2 \right)$
    \item $k_3^a = F^{\mathrm{FME}}_a\left(\braket{\hat a_p}^{\mathrm{FME}}(t) + k_2^a/2, \, \braket{\delta \hat a_i \delta \hat a_j}^{\mathrm{FME}}(t)+ k_2^{aa}/2, \, \braket{\delta \hat a^\dagger_i \delta \hat a_j}^{\mathrm{FME}}(t) + k_2^{ba}/2 \right)$
    \item $k_3^{aa} = F^{\mathrm{FME}}_{aa}\left(\braket{\hat a_p}^{\mathrm{FME}}(t) + k_2^a/2, \, \braket{\delta \hat a_i \delta \hat a_j}^{\mathrm{FME}}(t)+ k_2^{aa}/2, \, \braket{\delta \hat a^\dagger_i \delta \hat a_j}^{\mathrm{FME}}(t) + k_2^{ba}/2 \right)$
    \item $k_3^{ba} = F^{\mathrm{FME}}_{ba}\left(\braket{\hat a_p}^{\mathrm{FME}}(t) + k_2^a/2, \, \braket{\delta \hat a_i \delta \hat a_j}^{\mathrm{FME}}(t)+ k_2^{aa}/2, \, \braket{\delta \hat a^\dagger_i \delta \hat a_j}^{\mathrm{FME}}(t) + k_2^{ba}/2 \right)$
    \item $k_4^a = F^{\mathrm{FME}}_a\left(\braket{\hat a_p}^{\mathrm{FME}}(t) + k_3^a, \, \braket{\delta \hat a_i \delta \hat a_j}^{\mathrm{FME}}(t)+ k_3^{aa}, \, \braket{\delta \hat a^\dagger_i \delta \hat a_j}^{\mathrm{FME}}(t) + k_3^{ba} \right)$
    \item $k_4^{aa} = F^{\mathrm{FME}}_{aa}\left(\braket{\hat a_p}^{\mathrm{FME}}(t) + k_3^a, \, \braket{\delta \hat a_i \delta \hat a_j}^{\mathrm{FME}}(t)+ k_3^{aa}, \, \braket{\delta \hat a^\dagger_i \delta \hat a_j}^{\mathrm{FME}}(t) + k_3^{ba} \right)$
    \item $k_4^{ba} = F^{\mathrm{FME}}_{ba}\left(\braket{\hat a_p}^{\mathrm{FME}}(t) + k_3^a, \, \braket{\delta \hat a_i \delta \hat a_j}^{\mathrm{FME}}(t)+ k_3^{aa}, \, \braket{\delta \hat a^\dagger_i \delta \hat a_j}^{\mathrm{FME}}(t) + k_3^{ba} \right)$
    \item $\braket{\hat a_p}^{\mathrm{FME}}(t + \Delta t) = \frac{1}{6} \left( k^a_1 + 2k^a_2 + 2 k^a_3 + k^a_4 \right)$
    \item $\braket{\delta \hat a_i \delta \hat a_j}^{\mathrm{FME}}(t + \Delta t) = \frac{1}{6} \left( k^{aa}_1 + 2k^{aa}_2 + 2 k^{aa}_3 + k^{aa}_4 \right)$
    \item $\braket{\delta \hat a_i^\dagger \delta \hat a_j}^{\mathrm{FME}}(t + \Delta t) = \frac{1}{6} \left( k^{ba}_1 + 2k^{ba}_2 + 2 k^{ba}_3 + k^{ba}_4 \right)$
\end{enumerate}

which takes the field moments are time $t$, $\braket{\hat a}^{\mathrm{FME}}(t)$, $\braket{\delta \hat a_i \delta \hat a_j}^{\mathrm{FME}}(t)$, $\braket{\delta \hat a^\dagger_i \delta \hat a_j}^{\mathrm{FME}}(t)$, to a time $t + \Delta t$,  $\braket{\hat a}^{\mathrm{FME}}(t+\Delta t)$, $\braket{\delta \hat a_i \delta \hat a_j}^{\mathrm{FME}}(t+\Delta t)$, $\braket{\delta \hat a^\dagger_i \delta \hat a_j}^{\mathrm{FME}}(t+\Delta t)$. 
Where the functions are given by

\begin{align}
    g_1(a,\, \braket{\delta a_i \delta a_j} )_p &= \mathcal{F}\left[  \mathcal{F}^{-1} \left[ \mathcal{F}_y \left[ \braket{\delta \psi(x) \delta \psi(y) } \psi^\dagger (y) \right]_i \left( \frac{C}{k_i^2} + \Lambda_0 \right) \right] (x,x) \right]_p \\
    g_2(a,\, \braket{\delta a_i^\dagger \delta a_j} )_p &= \mathcal{F}\left[  \mathcal{F}^{-1} \left[ \mathcal{F}_x \left[ \braket{\delta \psi^\dagger(x) \delta \psi(y) } \psi (x) \right]_i \left( \frac{C}{k_i^2} + \Lambda_0 \right) \right] (x,x) \right]_p \nonumber \\
    &+\mathcal{F}\left[  \mathcal{F}^{-1} \left[ \mathcal{F} \left[ \braket{\delta \psi^\dagger(x) \delta \psi(x) } \right]_i \left( \frac{C}{k_i^2} + \Lambda_0 \right) \right] (x) \, \psi(x) \right]_p \\
    h(a)_{ij} &= \mathcal{F}_{xy} \left[ K(x,y) \, \psi(x) \, \psi(y) \right]_{ij} \\
    g_3(a,\, \braket{\delta a_i \delta a_j} )_{ij} &= \mathcal{F}_{xy} \left[  \mathcal{F}^{-1} \left[ \mathcal{F}_y \left[ \braket{\delta \psi(x) \delta \psi(y) } \psi^\dagger (x) \right]_i \left( \frac{C}{k_i^2} + \Lambda_0 \right) \right] (x,y) \psi (x) \right]_{ij} \nonumber \\
    &+ \mathcal{F}_{xy} \left[  \mathcal{F}^{-1} \left[ \mathcal{F} \left[ |\psi(x)|^2 \right]_i \left( \frac{C}{k_i^2} + \Lambda_0 \right) \right] (x) \, \braket{\delta \psi(x) \delta \psi(y) } \right]_{ij} + (i \leftrightarrow j) \\ 
    g_4(a,\, \braket{\delta a^\dagger_i \delta a_j} )_{ij} &= \mathcal{F}_{xy} \left[  \mathcal{F}^{-1} \left[ \mathcal{F}_x \left[ \braket{\delta \psi^\dagger(x) \delta \psi(y) } \psi (x) \right]_i \left( \frac{C}{k_i^2} + \Lambda_0 \right) \right] (x,y) \, \psi (x) \right]_{ij} + (i \leftrightarrow j) \\ 
    g_5 (a,\, \braket{\delta a_i \delta a_j} )_{ij} &= \mathcal{F}_{xy} \left[  \mathcal{F}^{-1} \left[ \mathcal{F}_x \left[ \braket{\delta \psi(x) \delta \psi(y) } \, \psi^\dagger (x) \right]_i \left( \frac{C}{k_i^2} + \Lambda_0 \right) \right] (x,y) \, \psi^\dagger (x) \right]_{ij} - (c.c., \,i \leftrightarrow j) \\ 
    g_6(a, \braket{\delta a^\dagger_i \delta a_j} )_{ij} &= \mathcal{F}_{xy} \left[  \mathcal{F}^{-1} \left[ \mathcal{F}_x \left[ \braket{\delta \psi^\dagger(x) \delta \psi(y) } \, \psi (x) \right]_i \left( \frac{C}{k_i^2} + \Lambda_0 \right) \right] (x,y) \, \psi^\dagger (x) \right]_{ij} \nonumber \\
    &+ \mathcal{F}_{xy} \left[  \mathcal{F}^{-1} \left[ \mathcal{F} \left[ |\psi(x)|^2 \right]_i \left( \frac{C}{k_i^2} + \Lambda_0 \right) \right] (x) \, \braket{\delta \psi^\dagger(x) \delta \psi(y) } \right]_{ij} - (c.c, \, i \leftrightarrow j) \,.
\end{align}
\end{widetext}
We again define our discrete Fourier transforms as in equation \eqref{paddFourier} in order to enforce non-periodic boundary conditions.
The position space fields are related to the momentum space arguments by 
\begin{align}
    \psi(x) &= \sum_i a_i u^\dagger_i(x) \, , \\
    \braket{\delta \psi^\dagger(x) \psi(y)} &= \sum_{ij} \braket{\delta \hat a^\dagger_i \delta \hat a_j} u_i(x) u_j^\dagger(y) \, , \\ 
    \braket{\delta \psi(x) \psi(y)} &= \sum_{ij} \braket{\delta \hat a_i \delta \hat a_j} u^\dagger_i(x) u_j^\dagger(y) \, .
\end{align}

\subsection{Quantum field theory}

The evolution of the quantum system is solved by integrating Schr\"odinger's equation

\begin{equation} \label{Schr}
    \partial_t \ket{\vec{z}} = -i \hat H \, \ket{\vec z} \,.
\end{equation}

We integrate this equation using the QIBS repository available at \href{https://github.com/andillio/QIBS}{https://github.com/andillio/QIBS}.

\section{Test problems} \label{sec:testProbs}

The purpose of this section is to compare the FME and MFT approximations of quantum systems using solutions which can be evaluated exactly, either analytically or numerically. In general, we will look to demonstrate that the field moment expansion is successful based on the following criteria

\begin{enumerate}
    \item Provides a more accurate approximation of the expectation value of the field operator, $\braket{\hat a}$ at least until the quantum breaktime as defined in equation \eqref{breakTime}. 
    \item An accurate approximation of when the quantum breaktime occurs.
\end{enumerate}

Note that the expectation values of the occupation numbers can be found using $Q$, $n_{tot}$, and $\braket{\hat a}$. Therefore, achieving the two goals listed above also implies that the field moment expansion can accurately approximate the expectation values of the occupation numbers.

\subsection{Kerr nonlinearity}

\begin{figure*}
	\includegraphics[width = .97\textwidth]{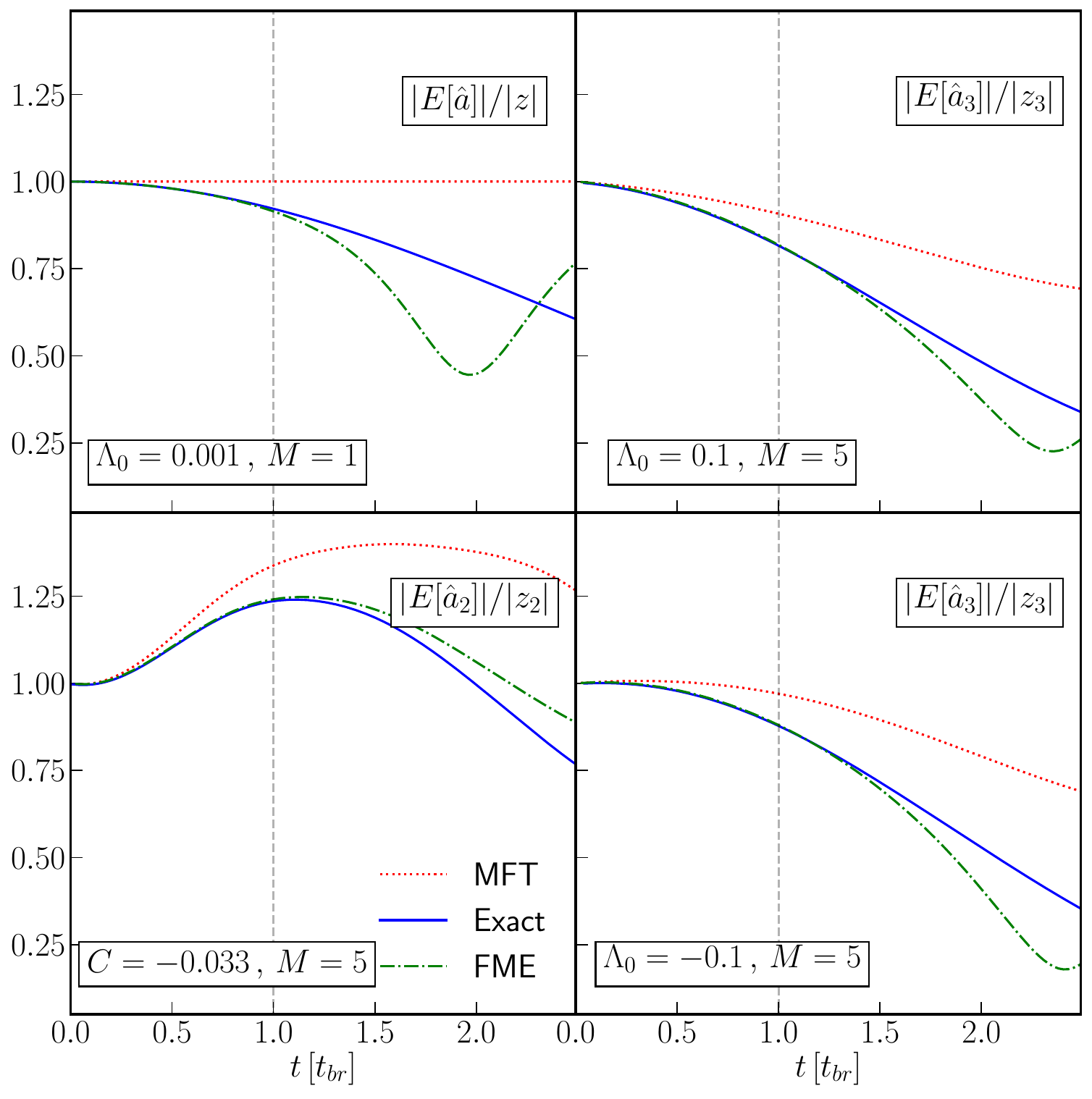}
	\caption{ Here we show the evolution of mean field for a given more for each method in each of our test problems. The Kerr nonlinearity is shown in the top left. The repulsive and attractive contact interactions in the top and bottom right respectively. Attractive long range interactions are shown in the bottom left. Nonzero coupling constants and mode numbers are given in each panel. On the vertical axis we plot the absolute value of the expectation of one of the mode field operators as a fraction of the initial value. On the horizontal axis we plot the time as a fraction of the quantum breaktime. Therefore, $t=1$ corresponds to the quantum breaktime in all plotted systems, shown in dashed lighted gray. In all cases, the classical field solution, shown in dotted red, has diverged from the exact quantum solution, shown in solid blue. However, the field moment expansion solution, shown in dashed green, remains an accurate approximation of the exact quantum solution at least until the breaktime.}
	\label{fig:compareAll}
\end{figure*}

\begin{figure*}
	\includegraphics[width = .99\textwidth]{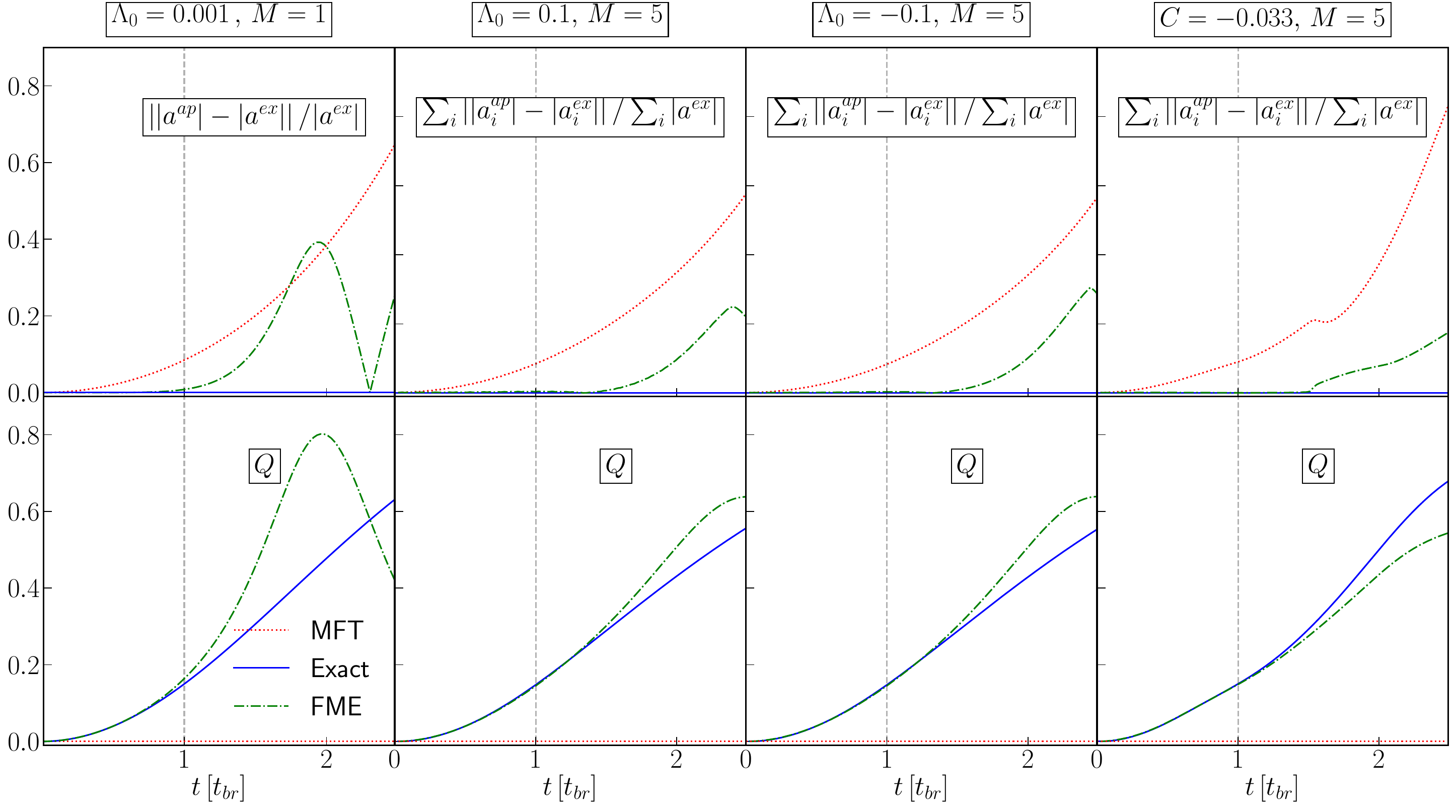}
	\caption{ Here we show the fractional error and approximation of $Q$ for each method in each of our test problems. In the top row we show the error in each method's approximation of the magnitude of the mean field as a fraction of the exact value of the mean field. In the bottom row we show each method's approximation of $Q$. The Kerr nonlinearity is shown on the far left column. The repulsive and attractive contact interactions in the middle left and right columns respectively. Attractive long range interactions are shown in the far right column. Nonzero coupling constants and mode numbers are given above each column. On the horizontal axis we plot the time as a fraction of the quantum breaktime. Therefore, $t=1$ corresponds to the quantum breaktime in all plotted systems, shown in dashed lighted gray. In all cases, the classical field, shown in dotted red, has a relatively large fractional error by the breaktime. However, the field moment expansion solution, shown in dashed green, remains an accurate approximation of the exact quantum solution, shown in blue, until at least the breaktime. Likewise, the field moment expansion accurately approximates $Q$ until the breaktime.}
	\label{fig:errQAll}
\end{figure*}

In this section we examine the Kerr nonlinearity which can be described by the following Hamiltonian

\begin{align}
    \hat H &= \omega \, \hat a^\dagger \hat a  + \frac{\Lambda_0}{2} (\hat{a}^\dagger \hat{a})^2 \nonumber \\
    &= \left( \omega + \frac{\Lambda_0}{2} \right) \, \hat a^\dagger \hat a  + \frac{\Lambda_0}{2} \hat{a}^\dagger \hat{a}^\dagger \hat{a} \hat{a} 
    \nonumber \\
    &= \omega_0 \, \hat a^\dagger \hat a  + \frac{\Lambda_0}{2} \hat{a}^\dagger \hat{a}^\dagger \hat{a} \hat{a} \, .
\end{align}

This is a special case of our Hamiltonian in equation \eqref{Ham} with $C=0,\, M=1$. 

This problem is interesting because it admits an exact solution, and so the time scales on which it diverges from the classical solution can be found analytically \cite{Yurke1986, TANAS1983351}. Our initial condition will be a coherent state, see equation \eqref{coherentStates}. The exact wavefunction can be given as a function of time as follows
\begin{equation}
    \ket{\phi (t)} = \exp \left[ -\frac{|z|^2}{2} \right] \sum_n e^{-it (\Lambda_0 n^2 /2  + \omega n)} \frac{z^n}{\sqrt{n!}} \ket{n} \, .
\end{equation}

\begin{figure}
	\includegraphics[width = .45\textwidth]{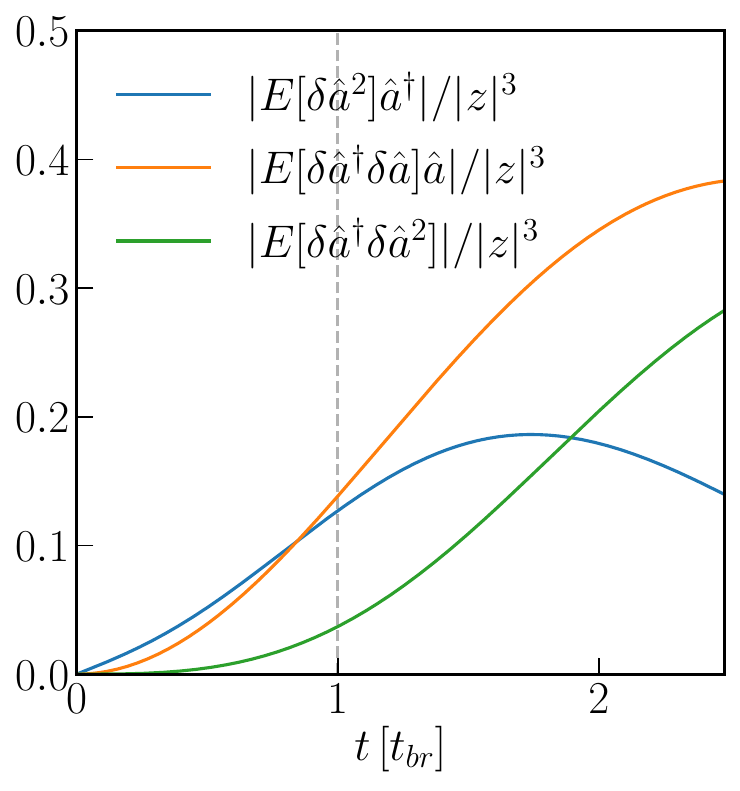}
	\caption{ Here we show the evolution of the central field moments, normalized by $|z|^3$, over time. We see all three moments become $\sim \mathcal{O}(1)$ on a time scale set by the nonlinearity. The second moments becoming relatively large by the quantum breaktime, shown in dashed light gray. The moments growth is hierarchical, i.e. the second moments become large before the third moment. Here we set $z = 5$, $\Lambda = 1 \times 10^{-3}$, and $\omega = 1$.}
	\label{fig:YurkeMoments}
\end{figure}

Given this, it is straightforward to calculate the exact evolution of the normally ordered central moments and the field expectation. The quantum evolution of the expectation of the field is characterized by a decaying amplitude which is not captured in the classical theory, as shown in the top left panel of figure \ref{fig:compareAll}. Here we set $z = 5$, $\Lambda =  1 \times 10^{-3}$, and $\omega = 1$.

The far left column of figure \ref{fig:errQAll} shows the result of applying the field moment expansion to this system.  We can see that $Q$ effectively parameterizes the time when the fractional error in the classical theory is no longer small. Until this point the field moment expansion provides a solution with a much lower fractional deviation. Likewise, until the breaktime the field moment expansion estimate of $Q$ remains accurate. Therefore, the field moment expansion provides both a more accurate solution until the breaktime, and successfully provides an accurate estimation of this time. 

The evolution central moments are shown in figure \ref{fig:YurkeMoments}.   As expected for an initially coherent state the central moments all start out at $0$ and then grow on a time scale set by the nonlinearity. We see also that the moments grow hierarchically, with the second central moments, normalized by $|z|^3$, becoming $\sim \mathcal{O}(1)$ faster than the third central moment. We see both the initial accuracy of the mean field theory and the hierarchically growth conditions are met for this system.

The field moment expansion remains more accurate, past the breaktime, however, we can see in figure \ref{fig:YurkeMoments} that past this time the third moment begins to become relevant. Since we have truncated our expansion at second order past this time is where we expect our solver to fail. Therefore, even assuming hierarchical growth, the field moment expansion is not reliable past the time when the highest moment in its truncation becomes large. 

\subsection{Contact interactions}

Unlike the previous test problem, scalar field dark matter systems involve a large number of modes. Therefore, it is prudent to test the accuracy of the field moment expansion on a system with multiple modes. We select the system given by the Hamiltonian in equation \eqref{Ham} with $M = 5$. This Hamiltonian has been used as a test problem in \cite{Hertberg2016, Sikivie:2016enz, Sikivie2012} and so will serve as a good benchmark to test the field moment expansion. It also contains a self-interaction term which is present in many models of scalar field dark matter.

This system models a contact interaction and linear dispersion, which implies the following

\begin{align} 
    \Lambda^{ij}_{pl} &= \Lambda_0 \, \delta^{ij}_{pl} \, , \\
    \omega_j &= j \,  \omega_0
\end{align}

Where $\Lambda_0 < 0$ defines an attractive interaction and $\Lambda_0 > 0$ a repulsive one. We evolve a coherent state defined by parameter $\vec z \in \mathbb{C}^5$. In order to test how the solution behaves with scaled occupation number we will simulate a benchmark coherent state $\ket{\vec z; r}$ where $\vec z = (0,\,\sqrt{2r}\, e^{i\theta_1},\,\sqrt{2r}\,e^{i\theta_2} ,\,\sqrt{1r}\,e^{i\theta_3},\,0)$ and the phases are drawn from a uniform random distribution, $\theta_i \sim U[0, 2\pi)$ with fixed random seed.

\begin{figure*}
	\includegraphics[width = .97\textwidth]{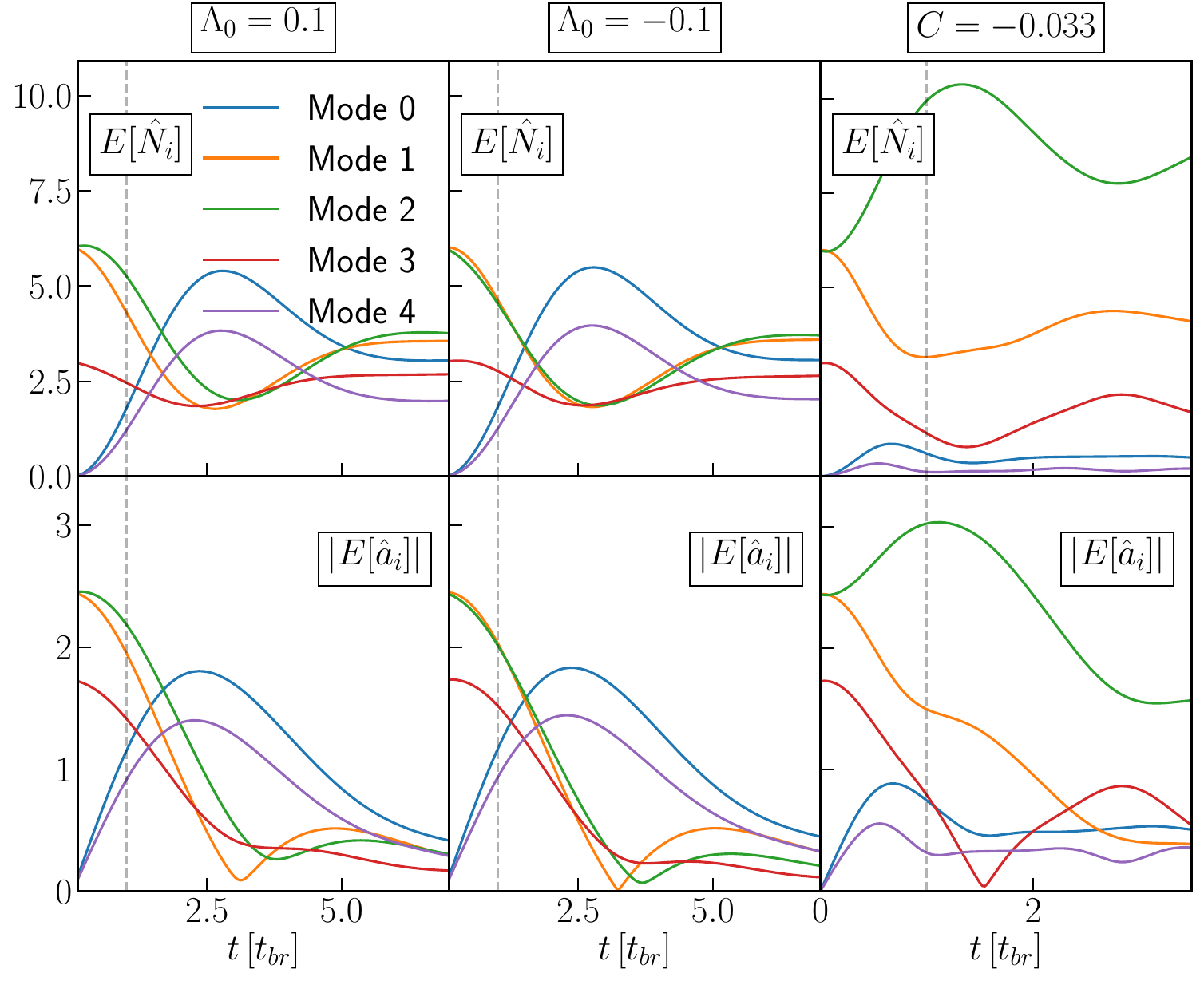}
	\caption{ Here we plot the evolution the expectation of the number and field operators. In the top row we show the expectation value of the occupation of each mode. In the bottom row we show the amplitude of the expectation of the field operator. In general this amplitude decays over time due to quantum effects like phase diffusion. The middle and left columns show the results of simulating contact interactions. The left column shows the repulsive case with $\Lambda_0 > 0$ and the middle column the attractive case with $\Lambda_0 < 0$. The right most column shows evolution of the long range attractive interactions. For reference, the quantum breaktime for each system is shown in dashed light gray. Nonzero nonlinear parameters are given above each column. In all cases $\omega_0 = 1$, $M=5$, and $r=3$. }
	\label{fig:allEvl}
\end{figure*}

For this system, in general, the occupations of the modes will thermalize and the expectation of the field itself will decay. This is shown in the right two columns of figure \ref{fig:allEvl}. Here we set $\Lambda_0 = \pm 0.1$, $\omega_0 = 1$, $M = 5$, and $r=3$.

We now test the field moment expansion to assess its accuracy and ability to approximate the quantum break time. As in the previous problem we can see that the field moment expansion solution remains close to the quantum solution past where the deviation between the classical field theory and quantum field theory becomes large. This is show for mode 3 in middle two columns of figure \ref{fig:compareAll}. 
\begin{figure}
	\includegraphics[width = .45\textwidth]{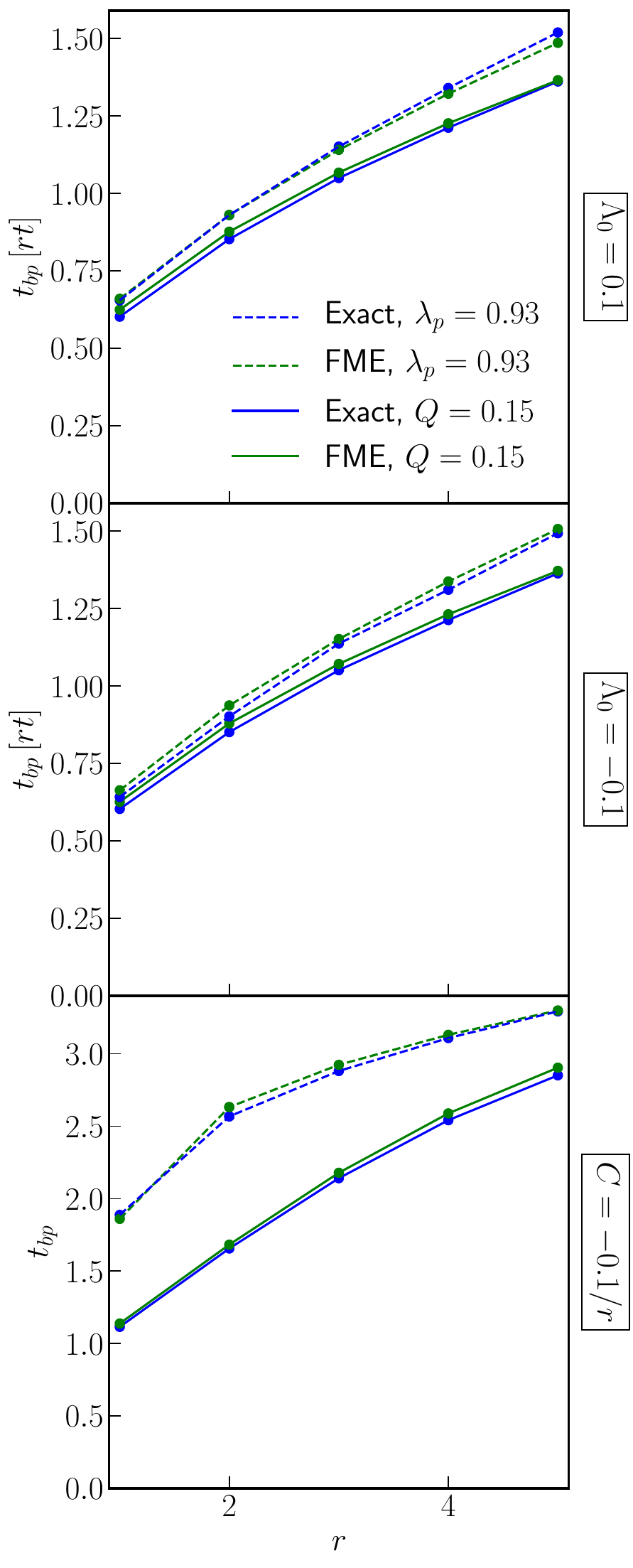}
	\caption{Here we show FME, in green, estimate of the breaktime compared to the exact quantum result, in blue, for a number of $r$ values. We show both the results of an analysis using $Q$ and the PO condition, dashed and solid, respectively. We see that FME provides a close estimate of the breaktime. Here we set $\omega_0 = 1$, $M = 5$.  }
	\label{fig:allBreaktimes}
\end{figure}
The field moment expansion also successfully predicts the quantum breaktime. This is shown for $r=3$ in the middle two columns of figure \ref{fig:errQAll}. We can see that the field moment expansion approximation of the field as well as the $Q$ parameter remains accurate until past the breaktime. We use the field moment expansion to estimate the breaktime for a number of different values of $r$. This is shown for the repulsive and attractive potentials in the top and middle panels, respectively, of figure \ref{fig:allBreaktimes}. There we also show an approximation of the breaktime using the PO condition. The results of the two breaktime definitions approximately agree. We see that the field moment expansion closely approximates the breaktime in all cases. 

\subsection{Long range interactions}

SFDM can include self-interactions like those in the previous section. However, given that we expect dark matter to be nearly collisionless, long range interactions like those found in gravity are going to govern much of the evolution. Therefore, we now turn towards modeling a system with a $1/r$ potential.

We still start with the Hamiltonian described in equation \eqref{Ham}. Like in contact interaction test problems we will use $M=5$. Long range interactions and a quadratic dispersion relation can be modeled using the following constants

\begin{align}
    \Lambda_{pl}^{ij} &= \left( \frac{C}{2(p_p - p_i)^2} + \frac{C}{2(p_p - p_j)^2} \right) \delta^{ij}_{pl} \, \\ 
    \omega_j &= \frac{j^2}{2} \omega_0 \, .
\end{align}

Where again $C<0$ gives an attractive potential, and $C>0$ a repulsive one. Here we evolve the same benchmark coherent state, $\ket{\vec z; r}$,  as the last section, where $\vec z = (0,\,2r\, e^{i\theta_1},\,2r\,e^{i\theta_2} ,\,1r\,e^{i\theta_3},\,0)$ and the phases are drawn from a uniform random distribution, $\theta_i \sim U[0, 2\pi)$ with fixed random seed.

The density of dark matter is well measured \cite{Planck}. And therefore the quantity $n_{tot} C$ should be fixed as we vary the occupation number. This means that larger occupation results in a lower value of $C$. As we scale our reference state then we will also scale the coupling constant sending $C \rightarrow C/r$. 

The evolution of this system is shown in the right column of figure \ref{fig:allEvl} for $r=3$. The field moment expansion produces an accurate estimation of the field until the breaktime, see the right most column of figure \ref{fig:compareAll}. We can see that the fractional error in the field moment expansion estimation of the field is close to zero up to and past the breaktime. Additionally, the field moment expansion successfully predicts $Q$ until the breaktime and consequently the breaktime itself, shown in the right most column of figure \ref{fig:errQAll}.

\section{Conclusions} \label{sec:conclusions}

In all the test problems the FME successfully approximated the first and second order moments of the exact quantum evolution when the correction terms were subleading order, i.e. when $Q \not\sim 1$. Therefore, we can say that the FME provides
\begin{enumerate}
    \item A more accurate approximation of the expectation value of the field operator, $\braket{\hat a}$ at least until the quantum breaktime as defined in equation \eqref{breakTime}. 
    \item An accurate approximation of when the quantum breaktime occurs.
\end{enumerate}
This is not terribly surprising because we solved the FME to second order. Intuitively, a second order approximation should remain accurate for longer than a first order approximation like MFT. Likewise, because the benchmarks of classicality we used, the $Q$ parameter and the PO condition, are themselves based on second order moments of the evolution, the fact that the FME can approximate the breaktime is not surprising. 

However, it is important to note that the method makes a number of assumptions about the system. Specifically, we make an assumption about the initial conditions and the evolution of the system as follows
\begin{enumerate}
    \item \textbf{Initial conditions assumption:} The initial conditions should be well approximated by the MFT. That is initially, $Q \ll 1$ and $1 - \lambda_p / n_{tot} \ll 1$. 
    \item \textbf{Evolution assumption:} The evolution of the field moments should be hierarchical. That is, for a term in the equations of motion of the field operators, $F$ of order $m$ written as a function of moments of order less than and equal to $m$, given $F^m(\text{moments of order $\le m$})$, over the evolution the terms must satisfy $F^1 > F^2 > F^3 \dots$.
\end{enumerate}
For coherent state initial conditions and the Hamiltonian in equation \eqref{Ham}, these assumptions are generally satisfied. However, for initial conditions such as number eigenstates, which do not have have hierarchically ordered moment terms, neither FME or MFT will accurately approximate the quantum evolution. The specific manner in which the quantum solutions approach a classical description will be explored in a later paper.

If it is known that the initial conditions of a system are well described by MFT, as in the case of coherent states, then the FME can provide a reliable check of the timescales on which the MFT approximation remains valid. This method could be applied to the evolution of ultra light scalar field dark matter created via the misalignment mechanism. Because the misalignment mechanism creates a system initially well described by the classical theory FME could be used to approximate the quantum breaktime of this system. This application will be explored in a later publication. 

In this work we have focused on the results of solvers that integrate equations expanded to second order in the field moments. The equations of motion are obtained via a truncation of an infinite series of coupled differential equations. However, it may be numerically feasible to instead create a closure relation using Wick's theorem, or as is done in \cite{Prezhdo2000}, which allows moments beyond second order to be approximated using lower order moments. Such a solver may remain accurate for longer than the current implementation. 

\begin{acknowledgments}
A.E., A.Z., and T.A. are supported by the U.S. Department of Energy under contract number DE-AC02-76SF00515.
\end{acknowledgments}

\bibliography{BIB}
\end{document}